# The "Horizon-T" Experiment: Extensive Air Showers Detection

[1] R.U. Beisembaev, E.A. Beisembaeva, O.D. Dalkarov, V.A. Ryabov, A.V. Stepanov, N.G. Vildanov, M.I. Vildanova, V.V. Zhukov

[2] K.A. Baigarin, D. Beznosko

[3] T.X. Sadykov

[4] N.S Suleymenov

**Abstract**

Horizon-T is an innovative detector system constructed to study Extensive Air Showers (EAS) in the energy range above $10^{16}$ eV coming from a wide range of zenith angles (0º - 85º). The system is located at Tien Shan high-altitude Science Station of Lebedev Physical Institute of the Russian Academy of Sciences at approximately 3340 meters above the sea level. It consists of eight charged particle detection points separated by the distance up to one kilometer as well as optical detector subsystem to view the Vavilov-Cherenkov light from the EAS.

The time resolution of charged particles and Vavilov-Cherenkov light photons passage of the detector system is a few ns. This level of resolution allows conducting research of atmospheric development of individual EAS.

## 1. Introduction

Tien Shan high-altitude Science Station, a part of Lebedev Physical Institute of the Russian Academy of Sciences, is located near the city of Almaty, KZ, at the altitude of about 3340 meters above sea level. The Horizon-T detector system has been constructed at the station for the purpose of EAS charged particles detection from the parent particle of energy above $10^{16}$ eV, and at the zenith angles from 0º to 85º. The system has been designed for the detailed study of spatial and temporal characteristics of the charged particles and Vavilov-Cherenkov photons distribution within each individual EAS.

[1] P. N. Lebedev Physical Institute of the Russian Academy of Sciences, Moscow, Russia
[2] Nazarbayev University, Astana, Kazakhstan
[3] LLP "Institute of Physics and Technology", Almaty, Kazakhstan
[4] Kazakh-British Technical University, Almaty, Kazakhstan

The Horizon-T detector system realizes ideas that were first formulated in [1], [2]. The first results of the Horizon-T are published in [3] and [4]. Latest information on the experiment status is available in [5] [6] [7], and for the latest result please see [8].

## 2. Spatial and Temporal EAS structure

As EAS develops while passing through the atmosphere, ultra-relativistic electrons, muons and Vavilov-Cherenkov radiation photons each form a so-called shower disk, a disk-like figure schematically drawn in Figure 1.

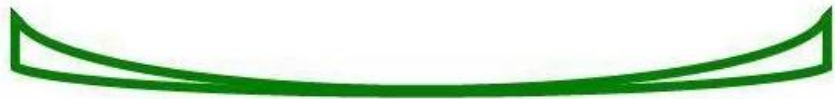

**Figure 1: Shower Disk relative shape.**

For EAS reaching to detector system at small angles to the vertical the diameters of all three disks can reach up to 1 km or so. Calculations, carried using CORSIKA [9] EAS simulation software package, indicate that for vertically incoming EAS with primary particle energy ~$10^{17}$ eV at the distance 100 m from the central axis the charged particles disk passes the observation level in ~20-30 ns. This sets a characteristic time scale for the individual detector needed to study the disk structure along its propagation axis at better than that scale, or better than about ~5ns to resolve any structure within the disk. Note that the time of passage grows to ~100-150 ns at 500 m from the EAS axis, thus allowing for lower time resolution on the order of ~20 ns.

With the increase of zenith angle the apparent thickness of the shower disks increases as well, thus the observation level passage time increases as well. The ability to conduct measurements at large zenith angles near to the horizon ($0^o$ - $85^o$) and the high time resolution are reflected in the name of the experiment: Horizon-T (where T stands for time).

## 3. Horizon-T Detector System Design

Time of passage of the charged particles disk through the setup is recorded using 8 detection points. The relative coordinates of every station and distances w.r.t. station 1 are presented in the Table 1. The bird's eye view of the setup and the positions of the points are shown in Figure 2.

The system center is indicated by a geodesic benchmark installed at the detection point 1 at the height of 3346.05 meters above the sea level and with geographical coordinates of 43°02′49.1532" N and 76°56′43.548" E. This benchmark is the origin for the XYZ coordinate system for Horizon-T. The X-axis is directed to the north, Y-axis to the west and Z-axis is directed vertically up.

**Table 1: Coordinates of detection points.**

| Station # | X, m | Y, m | Z, m | R, m |
|---|---|---|---|---|
| 1 | 0 | 0 | 0 | 0 |
| 2 | –445.9 | –85.6 | 2.8 | 454.1 |
| 3 | 384.9 | 79.5 | 36.1 | 394.7 |
| 4 | –55.0 | –94.0 | 31.1 | 113.3 |
| 5 | –142.4 | 36.9 | –12.6 | 147.6 |
| 6 | 151.2 | –17.9 | 31.3 | 155.4 |
| 7 | 88.6 | 178.4 | –39.0 | 203.0 |
| 8 | 221.3 | 262.0 | 160.7 | 378.7 |

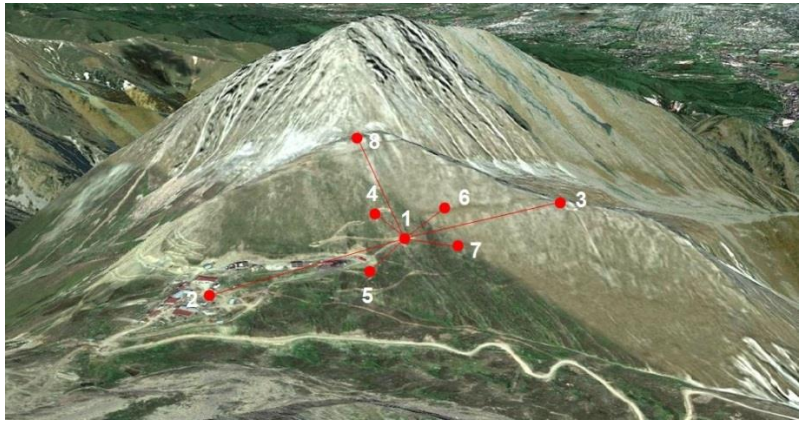

**Figure 2: The bird-eye view of the detector stations positions.**

### 3.1 Cherenkov Detector

The Vavilov-Cherenkov radiation detector (VCD), shown in Figure 3, is located next to detection point 1 as close to DAQ system as possible. It consists of three parabolic mirrors with 150 cm diameter and focal length of 65 cm each. They are mounted on the rotating support allowing detection of Vavilov-Cherenkov radiation in zenith angle range of $0°-80°$ and in azimuthal angle range of $0°-360°$. There is a MELZ [10] photoelectric multiplier tube (PMT) model PMT-49Б (FEU49B) and a Hamamatsu [11] H6527 PMT located in the focal point of two lower mirrors. Both are 15cm cathode diameter PMTs with the spectral response from 360 nm to 600 nm. The field of view of each mirror + PMT is ~$13°$.

From the geographical regions studied for the astroclimate, eastern Tien-Shan is well suited for Vavilov-Cherenkov radiation measurements [12] since for the most of the year there is a Rayleigh-type atmosphere when there is no aerosol present. This allows conducting the measurements of Vavilov-Cherenkov radiation from EAS in the large range of zenith angles up to $80°$. At the larger angles the station view of the horizon is limited by the surrounding mountaintops.

### 3.2 Scintillator Detectors

For the detection of charged particles, three scintillator detectors (SD) are located at each station and are oriented perpendicular to each other covering the x, y and z planes (Figure 4). The z-plane is parallel to the sky as mentioned before. This arrangement provides angular isotropy in the detection of charged particles incoming from the upper hemisphere. The mountains around the station do not interfere with EAS axis detection at angles near to the horizon, up to 85º.

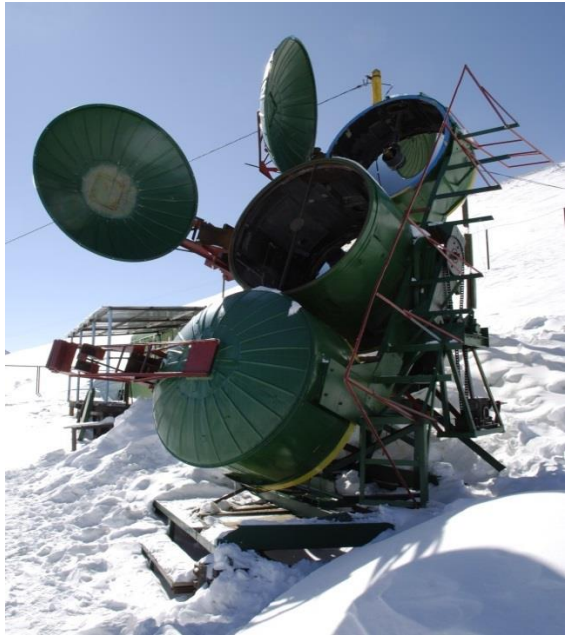

**Figure 3: The Vavilov-Cherenkov detector.**

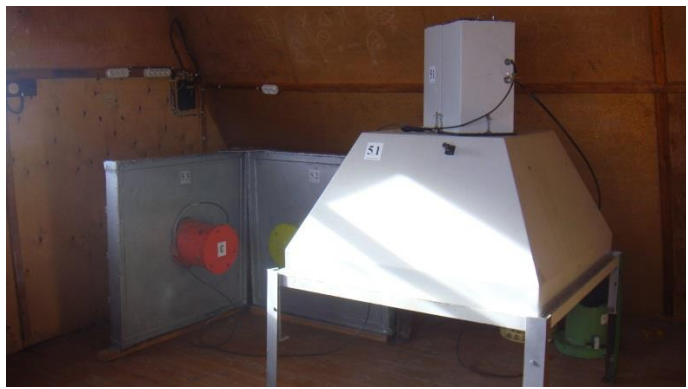

**Figure 4: Scintillator detectors orientation.**

A polystyrene-based square-shaped cast polystyrene scintillator with PPO fluor and POPOP shifter [13] with 1 m$^2$ area and 5 cm width is used in each SD. Light produced by a charged particle is registered by a PMT.

The biasing voltage on all detectors is adjusted such as the uniform particle detection efficiency is reached among the horizontal SD and among all vertically positioned detectors.

The PMT-49Б (FEU49B) and Hamamatsu R7723 PMTs are both used in scintillator detectors within the Horizon-T detector system. A typical pulse from the MIP (minimally ionizing particle) is used for each detector for characterization. The one of the characteristics discussed here is the duration and the uncertainty of the particle signal detection.

For analysis purposes, the pulse front and total duration are defined as following: the *pulse front* is between 0.1 and 0.5 of the total area under the pulse (e.g. of the pulse cumulative distribution, CDF) and the *total duration* is between the level 0.1 and level 0.9 of the pulse cumulative distribution. The example of a PMT pulse (inverted) and its normalized cumulative distribution are shown in Figure 5 (left) and in Figure 5 (right).

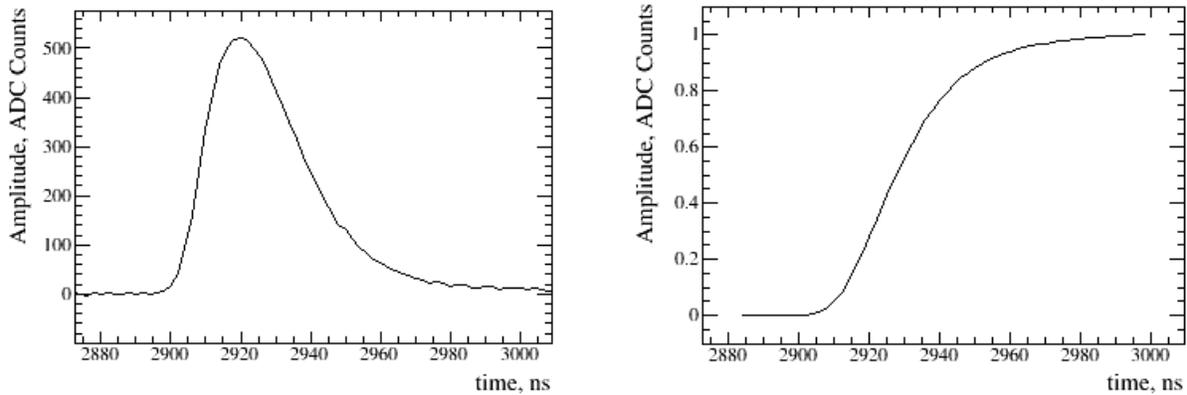

**Figure 5: Example of a typical inverted PMT pulse (left) and its normalized cumulative distribution (right)**

The time duration has been measured for the scintillator coupled both with the R7723 and the FEU49B PMTs. From the Figure 6 left and right sides we can see that the pulse front for the R7723 PMT with scintillator is 7.16±0.40 ns and the total duration is 21.6±1.48 ns with the systematic error of ±0.10 ns arising from the calculation of the pulse front and total duration.

The pulse front and the total duration for the FEU49B are shown in Figure 7 left and right sides and are 15.71±0.47 ns and 38.92±1.40 ns respectively with the ±0.09 ns systematic error.

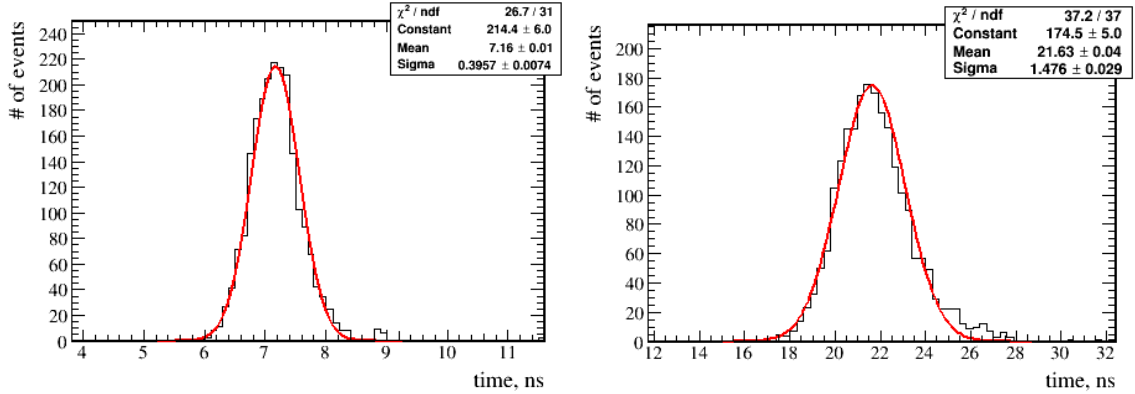

**Figure 6: Pulse front (left) and total duration for R7723 PMT with scintillator**

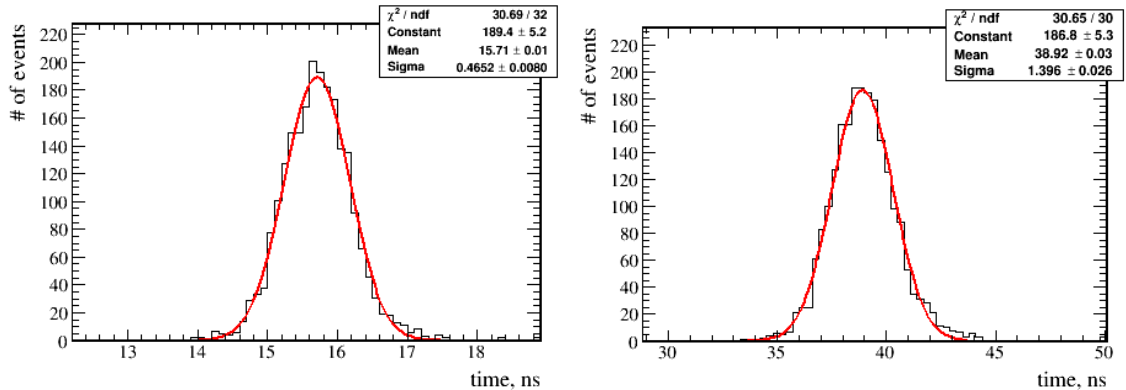

**Figure 7: Pulse front (left) and total duration for FEU49B PMT with scintillator**

### 3.3 Glass Detectors

In order to further increase the time resolution and to decouple the PMT and scintillator parts in pulse shaping, the glass-based detector with R7723 has been constructed [14] [15]. They are thoroughly calibrated as well [16]. The Vavilov-Cherenkov light from the charged particles in the glass is very fast compared to the PMT-induced pulse shape and is on the order of ~0.1 ns. From the Figure 8 left and right sides we can see that the pulse front for the R7723 PMT with glass for one MIP is 2.17±0.13 ns (over 18 m cable) and the total duration is 5.10±0.67 ns with the ±0.11 ns systematic error for both. These pulse characteristics correspond to the technical specs of the PMT.

Comparison of the results obtained using the scintillator and glass indicates that the scintillator contribution to the average pulse front duration about 5 ns, and its contribution to the total duration is about 15 ns. Because of its superior time resolution, the glass detector has been installed at the detection point #1 at the detector system center with work underway currently to also install glass detectors in detection points 4, 5, 6 and 7.

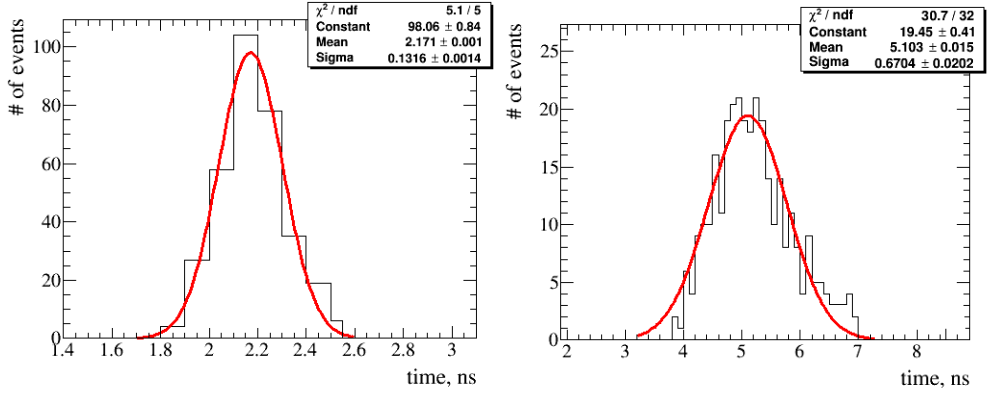

**Figure 8 : Pulse front (left) and total duration for R7723 PMT with glass**

## 4. Trigger Logic and Events Statistics

Each detection point sends two signal to the (DAQ) that is located at station 1, one from the horizontally installed (z-pane) SD, and one is a sum of the two vertical SD. All PMT signals are carried over the coaxial cables RK 75-7-316F-C SUPER produced by SpetsKabel [17] and impedance matched to the rest of the electronics and calibrated [18]. The system-wide electronic trigger is formed by a first 14-bit CAEN [19] DT5730 ADC (analog to digital converter) board. Three ADC boards (same model) in a common trigger schema make the data acquisition system (DAQ) system located immediately next to the detection point #1. The DAQ is triggered when the detection points (4 and 7) or (5 and 6), report the passage of charged particles from EAS disk. This relaxed hardware trigger allows keeping a larger data sample for further offline analysis. Typical offline trigger that is applied later requires a signal from all four detection points (4 & 5 & 6 & 7) and results in EAS registration intensity rate of ~7 events/hour at the Horizon-T.

The lowest energy at which Horizon-T starts to detect the EAS is denoted as threshold energy $E_{thr}$. By definition, the shower with $E_{thr}$ triggers the detector system if the EAS axis is between 0º and 30º zenith angle and passes between the trigger forming detector points (e.g. 4 and 7 or 5 and 6). Therefore, at the $E_{thr}$ the geometric factor for EAS detection at Horizon-T in the zenith angle range between 0º and 30º will be $\Gamma_{thr} = 0.06$ km² ster. Therefore, at the threshold energy $E_{thr}$ and with geometric factor $\Gamma_{thr}$ Horizon-T has the EAS detection rate of ~7 events/hour. From Figure 9, the all-particle cosmic-ray energy spectrum from [20], it can be seen that the corresponding energy to $E_{thr}$ is ~$10^{16}$ eV (marked by the left vertical red line).

As the energy of the primary particle increases, the charged particles density in the EAS increases, causing the widening of the zenith angle range and increasing the distance to the EAS axis that Horizon-T can detect. Thus, the geometric factor of the detector system increases. At the E=$10^{17}$ eV, the geometric factor $\Gamma(10^{17}$ eV$) = 1$ km² ster. From the energy spectrum of the primary particles (Figure 9) is follows that the event rate of the EAS from primaries of energies above $10^{17}$ eV should be about 1 event/hour.

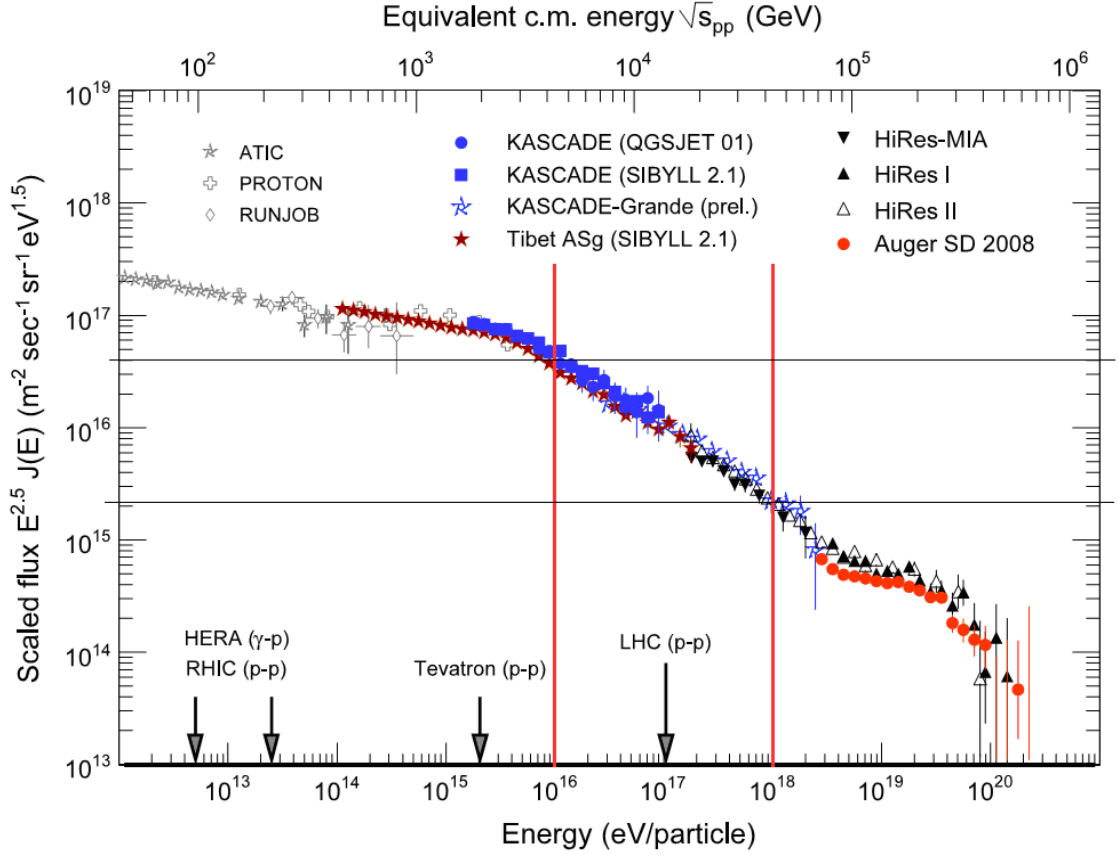

**Figure 9 : All-particle cosmic-ray energy spectrum [20]**

## 5. Conclusion

The Horizon-T detector system is designed for measuring the spatial and temporary structure of incoming EAS with axis in a wide range of zenith angles from 0º to 80º. Horizon-T is located at the altitude of about 3340 meters above sea level centered at geographical coordinates of 43°02′49" N and 76°56′43" E. The overall detection rate of EAS starting with primary particle energy at ~$10^{16}$ eV is 7 events/hour. For energies above $10^{17}$ eV, the rate is 1 event/hour.

Horizon-T has eight charged particle detection points with 24 scintillator detectors, each with 1 m² area. Detection points are separated by hundreds of meters.

SD detectors are read out by both FEU49B and Hamamatsu R7723 PMTs. The SDs with FEU49B have the front of the EAS disk detection resolution of 14.4±0.7 ns whereas SDs with R7723 have the disk front resolution of 6.1±0.9 ns.

The detector system also includes Vavilov- Cherenkov radiation detector that is made from three mirrors with FEU49B each that has the field of view of each mirror + PMT of ~13º. This detector allows measurements of light from EAS disks in the large range of zenith angles up to 80º. A detector with larger area and no long cables is planned as well [21], [22].

# Bibliography

[1]     R. U. Beisembaev, "Detector system project for studies of EAS at energies above 10^17 eV," *AS News, KZ Rep, phys-math series, #6 C,* pp. 74-76, 1991.

[2]     R.U. Beisembaev et al., "Muons of extra high energy horizontal EAS in geomagnetic field and nucleonic astronomy," *Proc. 24 ICRC. Roma. 1,* pp. 646-649, 1995.

[3]     R.U. Beisembaev et al., "Observations of the Cerenkov light and the charged particle of the EAS at large zenith angles," *Bul. Of RAS,* vol. 75 (3), p. 385, 2011.

[4]     R. U. Beisembaev at al., "The first results obtained with the HORIZON-T detector system.," *Journal of Physics,* vol. 409, 2013.

[5]     D. Beznosko et al., "Horizon-T Extensive Air Showers detector system operations and performance," in *PoS(ICHEP2016)784, proceedings of ICHEP2016*, Chicago, 2016.

[6]     D. Beznosko at al., "Horizon-T extensive air showers detector system operations and performance," *PoS Proceedings of Science,* PoS(ICHEP2016)784, 2016.

[7]     R. Beisembaev et al., "Horizon-T experiment status," in *EPJ Web Conf. 145 14001*, Moscow: ISVHECRI, 2017.

[8]     Rashid Beisembaev, Dmitriy Beznosko, Kanat Baigarin, Elena Beisembaeva, Oleg Dalkarov, Vladimir Ryabov, Turlan Sadykov, Sergei Shaulov, Aleksei Stepanov, Marina Vildanova, Nikolay Vildanov, Valeriy Zhukov, "Extensive Air Showers with unusual structure," in *EPJ Web Conf. 145 14001*, Moscow: ISVHECRI, 2017.

[9]     D. Heck, J. Knapp, J.N. Capdevielle, G. Schatz, T. Thouw, "CORSIKA: A Monte Carlo Code to Simulate Extensive Air Showers," *Forschungszentrum Karlsruhe Report FZKA,* vol. 6019, 1998.

[10]    MELZ-FEU, 4922-у pr-d, 4c5, Zelenograd, g. Moskva, Russia, 124482 (http://www.melz-feu.ru).

[11]     Hamamatsu Corporation, 360 Foothill Road, PO Box 6910, Bridgewater, NJ 08807-0919, USA; 314-5,Shimokanzo, Toyooka-village, Iwatagun,Shizuoka-ken, 438-0193 Japan.

[12]     T. B. Omarov, A. K. Kurchakov, B. I. Demchenko, U. M. Zavarzin., "Astroclimate of high altitude plateau Assi-Turgen," *Almaty: Science,* p. 59, 1982.

[13]     Adil Baitenov, Alexander Iakovlev, Dmitriy Beznosko, "Technical manual: a survey of scintillating medium for high-energy particle detection," *arXiv:1601.00086,* 2016/1/1.

[14]     M. Yessenov, A. Duspayev, T. Beremkulov, D. Beznosko, A. Iakovlev, M.I. Vildanova, K. Yelshibekov, V.V. Zhukov, "Glass-based charged particle detector performance for Horizon-T EAS detector system," *arxiv:1703.07919,* 03/2017.

[15]     R.U. Beisembaev, D. Beznosko, E.A. Beisembaeva, A. Duspayev, A. Iakovlev, T.X. Sadykov, T. Uakhitov, M.I. Vildanova, M. Yessenov and V.V. Zhukov, "Fast and simple glass-based charged particles detector with large linear detection range," *Journal of Instrumentation,* vol. 12, no. T07008, 2017.

[16]     D. Beznosko, T. Beremkulov, A. Iakovlev, A. Duspayev, M. I. Vildanova, T. Uakhitov, K. Yelshibekov, M. Yessenov, V.V. Zhukov, "Horizon-T Experiment Calibrations – MIP Signal from Scintillator and Glass Detectors," *arxiv:1703.07559,* 3/2017.

[17]     SpetsKabel Inc., 6/1-5 Birusinka St., Moscow, Russia. http://www.spetskabel.ru/.

[18]     D Beznosko, T Beremkulov, A Iakovlev, Z Makhataeva, M I Vildanova, K Yelshibekov, VV Zhukov, "Horizon-T Experiment Calibrations-Cables," *arXiv:1608.04312,* 8/2016.

[19]     CAEN S.p.A. Via della Vetraia, 11, 55049 Viareggio Lucca, Italy. http://caen.it..

[20]     Blumer J. et al., "Cosmic rays from the knee to the highest energiees.," *Progress in Particle and Nuclear Physics,* vol. 63, pp. 293-338, 2009.

[21]     Duspayev et al., "The distributed particle detectors and data acquisition modules for Extensive Air Shower measurements at "Horizon-T KZ" experiment,"

in *PoS(PhotoDet2015)056, in proceedings to PhotoDet2015 conference*, Moscow, 2015.

[22]    Duspayev, A., R. U. Beisembaev, T. Beremkulov, D. Beznosko, A. Iakovlev, K. Yelshibekov, M. Yessenov, V. Zhukov, "Simulation, design and testing of the HT-KZ Ultra-high energy cosmic rays detector system," *Proceedings of ICHEP2016,* vol. PoS(ICHEP2016)721, 2016.